\documentclass[letterpaper,12pt]{article}
\usepackage[utf8]{inputenc}
\usepackage[T1]{fontenc}
\usepackage{geometry}
\usepackage[doublespacing]{setspace}
\usepackage{graphicx}
\usepackage{natbib}
\bibpunct[: ]{(}{)}{;}{a}{}{,}
\usepackage{amssymb}
\usepackage{amsmath}
\usepackage{dcolumn}
\usepackage{booktabs}
\usepackage{ntheorem}
\newtheorem{hypothesis}{Hypothesis}
\usepackage{hyperref}[pdfpagelabels,bookmarks]
\hypersetup{
	pdftitle={},
	pdfauthor={},
	pdfsubject={},
	pdfkeywords={},
	breaklinks=true,
	colorlinks=true,
	urlcolor=black,
	linkcolor=black,
	citecolor=black,
	bookmarksnumbered
}

\title{Endogenous Coalition Formation in Policy Debates}
\author{Philip Leifeld\footnote{University of Manchester, Department of Social Statistics, \href{mailto:philip.leifeld@manchester.ac.uk}{\texttt{philip.leifeld@manchester.ac.uk}}.} \and Laurence Brandenberger\footnote{ETH Zurich, Chair of Systems Design, \href{mailto:lbrandenberger@ethz.ch}{\texttt{lbrandenberger@ethz.ch}} and University of Zurich, Institute of Political Science.}}
\date{This version: 23 May 2024}

\begin{document}

\pagenumbering{gobble}

\maketitle

\begin{abstract}
	\noindent Political actors form coalitions around their joint normative beliefs in order to influence the policy process on contentious issues such as climate change or population ageing.
	Policy process theory maintains that learning within and across coalitions is a central predictor of coalition formation and policy change but has yet to explain how policy learning works.
	The present article explains the formation and maintenance of coalitions by focusing on the ways actors adopt policy beliefs from other actors in policy debates.
	A policy debate is a complex social system in which temporal network dependence guides how actors contribute ideological statements to the debate.
	Belief adoption matters in three complementary ways: \emph{bonding}, which exploits cues within coalitions; \emph{bridging}, which explores new beliefs outside one's perimeter in the debate; and \emph{repulsion}, which reinforces polarization between coalitions and cements their belief systems.
	We formalize this theory of endogenous coalition formation in policy debates and test it on a micro-level empirical dataset using statistical network analysis and event history analysis.
\end{abstract}

\newpage

\pagenumbering{arabic}

\section{Introduction}
Policy process theory and work on lobbying suggest that political actors form coalitions around their joint beliefs in order to influence the policy process \citep{hajer1995politics, henry2011belief, hojnacki1997interest, junk2019diversity, leifeld2013reconceptualizing, sabatier1993policy, sabatier2014theories, weible2009themes}.
Despite a plethora of contributions on the existence and significance of ``advocacy coalitions'' and ``discourse coalitions'', little systematic evidence has yet been presented on how coalitions are formed and maintained over time around joint policy beliefs.

The literature argues that \emph{learning} plays a key role in the formation and maintenance of coalitions \citep[e.\,g.,][]{moyson2017cognition, sabatier1993policy, weible2023advocacy}, but the mechanisms through which learning is connected to coalition structure largely remain a black box.
Instead, existing work focuses on belief coalitions as an independent variable and coordination between actors as an explanandum \citep[e.\,g.,][]{henry2011belief}.
While this development helps explain the structure of policy networks as a function of actors' policy beliefs, the original puzzle of belief formation and, thereby, coalition formation around shared beliefs persists.
In our contribution, we therefore unpack the notion of policy learning within and across coalitions by postulating and testing three ways through which learning matters for coalitions: learning through bonding, bridging, and repulsion.

First, as actors in a complex policy process face substantial uncertainty and risk over adopting the right policy beliefs, they learn from other actors by utilizing \emph{bonding} relationships.
Bonding in this context means that actors infer who their adjacent coalition members are by observing the policy debate, and they learn from those actors who hold strongly similar beliefs.
Bonding in policy debates serves as a heuristic to minimize the risk of adopting the wrong policy beliefs.
This extends the ``risk hypothesis'' from the literature on collaborative governance and policy networks \citep{berardo2010self, berardo2014bridging, berardo2016understanding} to the case of belief formation and coalition formation.
We provide micro-level evidence for the role of bonding over beliefs in shaping coalition formation.

Second, in addition to these strong belief ties, actors learn beliefs through weak ties.
For this to happen, two conditions must be satisfied:
If an actor observes that many other actors expressed a certain policy belief in the recent policy debate \emph{and} if many of these other actors had at least a minimal degree of belief congruence with the focal actor around other issues, this signals to the actor that the policy belief in question is ideologically compatible and worth adopting.
Thus, actors not only seek strong bonding relationships; they also actively seek to adopt innovative beliefs that are compatible with their remaining preference profile through \emph{bridging} relationships.
The notion of bridging is known from the literature on collaborative governance, where actors form coordination ties with actors outside their trusted circles to explore opportunities in the wider policy network \citep{berardo2010self, berardo2014bridging, berardo2016understanding}.
Applied to the context of policy debates, learning through bridging serves to explore candidate beliefs held by the wider network of actors, potentially in an opposing coalition, while requiring at least minimal past belief congruence with those from whom these beliefs are adopted.
The distinction between bonding and bridging speaks to the often-documented decision dilemma between exploitation and exploration \citep{cohen2007should}:
While bonding embodies exploitation, or a search in the local belief neighbourhood, bridging embodies exploration by considering potentially compatible beliefs outside one's coalition.

Third, learning entails drawing experiences from negative examples.
Actors consider with whom they had substantive conflicts in the past, and they tend to oppose the current policy beliefs of these actors.
This reinforcement of conflictual relationships leads to repulsion between different coalitions and, therefore, complements bonding, which serves to increase the congruence within coalitions.
Negative relations, over and above the absence of relations, have been shown to be a constituent part of political polarization, for example in the US Congress \citep{neal2020sign}.
In policy debates, past belief conflicts spur new belief conflicts, which leads to the macro outcome of polarization.
Repulsion can co-exist with bridging across coalitions because bridging requires past minimal compatibility for learning while repulsion builds on past conflict.

The three mechanisms together are complementary.
They serve to increase congruence within coalitions through mutual reinforcement of positive relationships, learning of belief innovations across coalitions, and repulsion between coalitions through reinforcement of conflictual relationships.
We test this theory of coalition formation and belief learning using dynamic inferential network analysis and an empirical policy process over the course of nine years at a daily time resolution.
We find micro-level evidence that supports our three-pronged theory of endogenous coalition formation.

\section{How Policy Debates Unfold over Time}
Politics is about ``who gets what, when, how'' \citep{lasswell1950politics}.
This extends far beyond institutional arenas like Congress or the executive.
How political issues are understood and why some solutions to political problems make it onto the parliamentary and executive agendas while others are weeded out during public debate is at least of equal importance but lacks systematic understanding \citep{downs1972up}.
Research on these early stages of the policy process \citep{sabatier2014theories} has been struggling to provide convincing answers to one of the most pressing questions on agenda setting:
How do political actors, such as interest groups or legislators, decide to contribute normative statements about their desired policy outcomes to the public debate?
In other words, how do policy debates operate at the micro-level?

\begin{figure}
  \centering
  \includegraphics[width=0.6\textwidth]{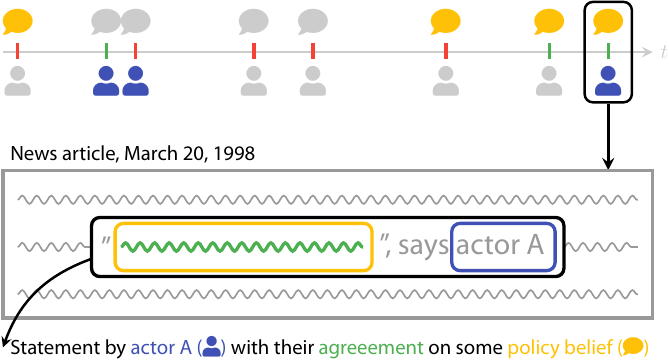}
  \caption{Illustration of policy debates as an event sequence of statements in news media articles.
           Each news article contains a number of text portions in which an actor expresses a policy belief, either by supporting or rejecting the belief.
           Different actors, beliefs, and signs are represented by colors.
           The actor--belief--agreement statement tuples can be put on a timeline, which forms an event sequence for statistical analysis.}
  \label{fig:dna}
\end{figure}

Most newspaper articles on the politics of the day contain quotes by political actors advocating or rejecting certain policy instruments or proposals (Figure~\ref{fig:dna}).
They are often in direct support or contradiction of previous statements by other political actors.
Lay readers, journalists, and political analysts try to make sense of these statements by inferring ``camps'', ``advocacy coalitions'', or ``discourse coalitions'' from a policy debate \citep{sabatier1993policy, hajer1995politics}.
In many cases, the cleavage lines seem clear, either because it is easy to associate interest groups with the policies they lobby for or because policies appeal in obvious ways to partisan ideologies of actors who run for office or a seat.
More often, however, it is unclear why an actor adopts a specific issue stance and why this happens at the specific time it is observed.
The prevailing view in the literature is that policy preferences are fixed and allow only for minor modifications through policy learning \citep{sabatier1993policy, hajer1995politics}.
Yet, the mechanisms through which learning occurs are opaque and need to be unpacked and tested.

Students of political processes are often content with describing macro-level variation, such as the presence of multiple advocacy or discourse coalitions in policy processes (as an example of variation across actors and within a time unit), or issue attention cycles with varying levels of salience of issues over time (as an example of temporal variation without any cross-sectional differentiation between actors).
We argue that such macro-level phenomena can be traced back to micro-level mechanisms in a theoretically meaningful way.
These mechanisms can be disentangled by considering and explaining individual statements of political actors and their timing.

However, individual statements about actors' policy beliefs or stances are rarely independent of each other.
Actors do not argue in isolation.
The very essence of a political debate is that actors interact with each other in order to ``influence the political discourse of their day'' \citep{hall1993policy}, both collaboratively and in conflict.
To see this, consider the counterfactual that no political discussion would be going on among political actors if it were impossible to influence other actors by making one's voice heard through the media and other outlets, possibly by altering public opinion (``issue campaigning'') or convincing other actors more directly (``policy learning'').
Hence, while useful for some purposes, we must overcome the prevailing view that policy preferences are fixed and instead devise empirical strategies to analyze how learning takes place.
We need to understand how these interdependencies shape the complex structure of political debates that ultimately leads to the emergence of coalitions in the policy process.

Our contribution proposes a set of theoretical mechanisms at the micro-level that lead endogenously to the formation and maintenance of coalitions in policy debates.
When actors decide to make a statement in the debate, they consider the past event sequence $E_{t - 1} = \left\{e_1, \ldots, e_{t - 1}\right\}$ of statements that were made before the current time $t$ (Figure~\ref{fig:dna}).
The past event sequence contains events $e = \left(s_e, r_e, a_e, t_e\right)$, comprising event time $t_e$, a sender $s_e \in S$ (the actor), a receiver $r_e \in R$ (the policy belief), and a relationship type $a_e \in A$ (support or rejection).
The probability of which statement event happens at $t$ is influenced by functions of the past event sequence and can be viewed from the perspective of each actor deciding between all possible belief--agreement combinations.

The set of actors $S$ includes various interest groups, legislators, and other actors who have a potential interest in expressing their opinion on the policy topic of the debate.
The set of policy beliefs $R$ contains specific policy instruments or proposals.
For example, in the debate on how to design or change a public pension system to address the financial pressure caused by population aging, policy beliefs can be the various policy proposals used in the debate, such as ``increase contribution level'', ``subsidize private pension schemes with taxes to complement the public pension system'', or ``tie pension level to the number of children of the person''.
They are specific and cannot be reduced to more fine-grained sub-beliefs other than on an ordinal scale, as in more or less of a specific instrument.

In considering the past event sequence, actors place a higher weight on recent statements and a lower weight on statements that occurred in the more distant past.
We assume that actors down-weight past statements by an exponential temporal decay function first introduced in the literature on relational event models of international conflicts \citep{brandes2009networks, lerner2013modeling}, with a half-life parameter of $T_{1/2} = 20$ days (though the results are relatively robust to the exact specification of $T_{1/2}$):
\begin{equation}\label{eq:decay}
	w(t_e, t, T_{1/2}) = \exp \left\{-(t - t_e) \left( \frac{ln(2)}{T_{1/2}} \right) \right\} \frac{ln(2)}{T_{1/2}},
\end{equation}
where $t$ is the current time point, $t_e$ is a prior time point, and half-life parameter $T_{1/2}$ determines after how many days the weight of the statement is halved.

\section{Coalitions in the Policy Process}
Coalitions are a central element in theories of the policy process \citep{sabatier2014theories}.
The Advocacy Coalition Framework (ACF) attributes a central role to clusters of actors with congruent policy beliefs \citep{sabatier1993policy}.
According to proponents of the ACF, there are usually between two and five coalitions in any given policy process.

Advocacy coalitions emerge because of overlapping functional roles of actors \citep{zafonte1998shared} and, primarily, because of mutually reinforcing policy learning within coalitions \citep{sabatier1993policy}, thus leading to a remarkable stability of coalitions over multiple decades.
In a policy network of actors with partly compatible and partly incompatible policy preferences, actors with similar institutional roles will stick together and mutually reinforce and complement each other's positions.
Two non-governmental organizations, for example, may initially hold only partly congruent policy beliefs but are likely to adopt and reinforce each other's policy beliefs, rather than their opponents' beliefs.
They have an incentive to do so in order to gain or maintain a position of power in the policy network through the operation of a cohesive coalition with coherent policy beliefs.

However, coalitions need not rely on institutional cues like overlapping functional roles for policy instrument learning.
Coalitions can also emerge as the result of previous states of the network, as in a Markov process.
For example, if actors $A$ and $B$ demonstrated relatively congruent policy preferences in the recent past and actor $C$ demonstrated a relatively dissimilar preference profile, then policy beliefs may diffuse from $A$ to $B$ or vice-versa.
Such a diffusion of policy beliefs among actors is the result of perceived joint coalition membership---without the actual need for a formal shared institutional role.
Learning is rather the result of each actor's observation of the other actors' stated beliefs.
This leads to increasing homogeneity of policy beliefs within coalitions over time \citep[see also][]{leifeld2014polarization}.
We call coalition formation dynamics that do not involve exogenous covariates, such as an actor's institutional role, \emph{endogenous} mechanisms of coalition formation.
Both types of coalition formation, endogenous and exogenous, can interact in a single policy debate.

The literature on opinion dynamics has explored similar dynamics of endogenous coalition formation \citep[e.\,g.,][]{altafini2012dynamics, antal2005dynamics, marvel2011continuous, quattrociocchi2014opinion}, where nodes in a network strive for structural balance in opinions in their local network environment \citep{cartwright1956structural}.
An actor with friends who hold similar opinions will be reinforced in his or her views while different opinions in the network neighborhood may lead to a tipping of the actor's opinion.
The extreme outcome of a completely balanced system, in which the network breaks apart into two entirely separate but internally homogenous coalitions, is known as a monotone dynamical system \citep{altafini2012dynamics} and exhibits complete opinion polarization.
Empirical networks, however, usually feature a mix of balanced and unbalanced triads because the endogenous coalition formation mechanism through learning and reinforcement within coalitions is complemented by other theoretical mechanisms.

The popular notion of ``discourse coalitions'' in the literature on argumentative discourse analysis \citep{hajer1995politics} captures essentially the same dynamic as posited by the ACF.
The basic premise is that actors try to manipulate other actors' view of reality through the articulation of arguments in a policy debate.
The success of a discourse coalition in manipulating a debate is a function of the coherence of the coalition's set of arguments \citep{leifeld2012political}.
Hence, like in the ACF, actors with already similar positions have an incentive to further mutually reinforce and adopt each other's statements in a policy debate in order to appear more convincing and exercise greater power.

In other words, actors learn from other actors in a policy debate, and the outcome of these learning processes is coalitions.
This learning of policy beliefs can occur through multiple plausible mechanisms.
In the following paragraphs, we seek to unpack these mechanisms, which contribute to the formation and maintenance of coalitions in a policy debate: bonding, bridging, and repulsion.

\section{Policy Learning within and across Coalitions} 

\subsection{Learning through Bonding}
In collaborative governance and policy networks, the risk hypothesis states that actors establish transitive and reciprocal ties (``bonding'') to tighten the local network structure around them when actors are at risk of defection \citep{berardo2010self, berardo2014bridging, berardo2016understanding}.
In the context of policy debates, actors also face a risk, though a different one.
Policy beliefs may be partly intransitive and incongruent, which may create uncertainty over which policy beliefs to adopt.
For example, if belief $r_1$ is incompatible with $r_2$ and $r_2$ is incompatible with $r_3$ and an actor holds belief $r_1$, should the actor adopt also adopt $r_3$?
If an actor has supported $r_1$ and rejected $r_2$ and it emerges that a new belief $r_3$ is incompatible with both $r_1$ and $r_2$, should the actor support or reject $r_3$?
This uncertainty over which policy beliefs to hold in one's portfolio and express in the policy debate creates the need for considering heuristics over the past event sequence.

Faced with the risk of adopting, and publicly repeating the wrong policy belief, actors turn to bonding.
Bonding in policy debates means an actor relies on cues from very congruent other actors who have expressed the same agreement pattern over many other policy beliefs in the past event sequence.
Bonding minimizes the actor's risk by trusting the assessment by actors who have so far displayed judgment that is very much in line with the actor's own judgment.
Learning from these actors serves to minimize uncertainty and risk by relying on trustworthy information.
The odds of being attacked by the opponent for holding incongruent views or a double standard are low if other actors with a similar coalition position advanced similar claims in the past.

\begin{figure}
	\centering
	\includegraphics[width=0.6\textwidth]{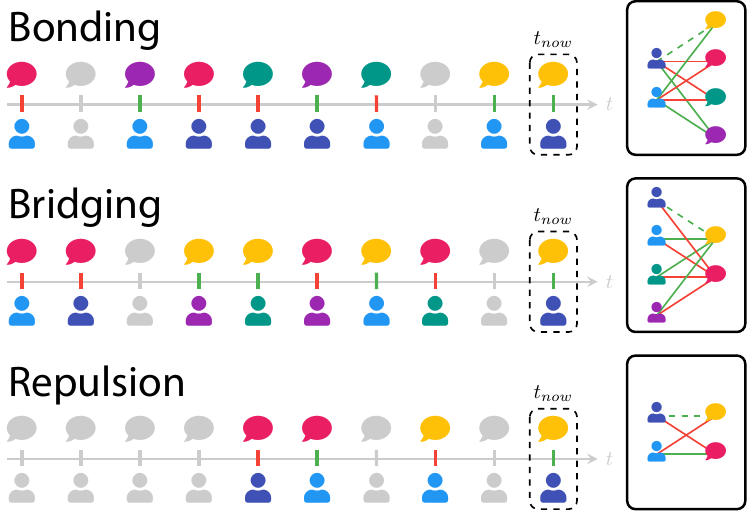}
	\caption{Illustration of three mechanisms of endogenous coalition formation. \emph{Bonding}: An actor has a higher probability to adopt a specific policy belief with a specific agreement pattern if another actor who shared many belief--agreement combinations in the past event sequence also adopted the same belief with the same agreement pattern. \emph{Bridging}: An actor has a higher probability of adopting a belief if many other actors with whom the actor shared at least minimal overlap in belief--agreement combinations in the past also adopted the same belief with the same agreement pattern. \emph{Repulsion}: An actor is more likely to adopt a belief with a specific agreement pattern if the belief was previously adopted with a conflicting agreement pattern by many other actors who previously adopted conflicting belief--agreement combinations in the past event sequence. The three mechanisms together shape the formation and maintenance of coalitions.}
	\label{fig:mechanisms}
\end{figure}

We operationalize bonding in policy debates using a network statistic, as shown in the upper event sequence in Figure~\ref{fig:mechanisms}.
The statistic counts (temporally weighted) balanced four-cycles.
In a signed bipartite graph, an actor $s_1$ will tend to make a positive statement about policy belief $r_1$ if other actors $s_i$ also referred to $r_1$ in a positive way and these other actors already shared many other beliefs $r_j$ with $s_1$ with the same agreement pattern (i.\,e., both positive or both negative regarding $r_j$).
Similarly, $s_1$ will tend to make a negative statement about $r_1$ if $s_1$ shared beliefs $r_j$ with other actors $s_i$ with the same agreement pattern (both positive or both negative) in the past and if these other actors $s_j$ also refer to $r_1$ in a negative way.
This yields many possible combinations of balanced four-cycles.
All of these cases reflect the tendency of an actor to adopt a policy belief from other actors with a similar expressed belief portfolio.
They can be counted over the past event sequence and aggregated to create a network statistic that operationalizes bonding.

At the same time, it matters for policy learning within coalitions \emph{when} these prior statements were observed by focal actor $s_1$.
If they all occurred in the recent past, the mutual support and learning is stronger because the focal actor has a good memory of the structure of the debate and these earlier statements, therefore, act as a stronger signal to $s_1$.
Conversely, trigger statements in the distant past, say, five years ago, may have a weak influence on $s_1$'s discursive activity.
The statistic takes this temporal decay into account by applying the decay function in Equation~\ref{eq:decay} to each past event:

\begin{equation}
	\begin{split}
		u_\text{bonding}(e, E_{t-1}, T_{1/2}) = & \sum_{e_i \in E_{t-1}} \sum_{e_j \in E_{t-1}} \sum_{e_k \in E_{t-1}} \bigg\{ [s_{e_i} \neq s_e] [r_{e_i} = r_e] [a_{e_i} = a_e] [s_{e_j} = s_e] \\
		& [r_{e_j} \neq r_e] [s_{e_k} = s_e] [r_{e_k} \neq r_e] [a_{e_j} = a_{e_k}] [t_{e_i} < t_{e_j}] [t_{e_j} < t_{e_k}] \\
		& \sqrt[3]{ w(t_{e_i}, t_e, T_{1/2}) w(t_{e_j}, t_e, T_{1/2}) w(t_{e_k}, t_e, T_{1/2}) } \bigg\},
	\end{split}
\end{equation}
where $e_i, e_j, e_k$ denote different statement events from the past event sequence and $[\ldots]$ are Iverson brackets, which return 1 if the condition in the square brackets is true and 0 otherwise.

The bonding statistic iterates over three statements in the past event sequence with respect to the current statement candidate $e$.
The chain of Iverson brackets before the cubic root indicates whether the three statements form a balanced four-cycle with $e$, indicating positive reciprocity between the focal actor $s_e$ and another actor.
The first three conditions correspond to the green tie from the light blue actor to the yellow belief in the bonding illustration in Figure~\ref{fig:mechanisms}.
The next two conditions correspond to a tie from the dark blue actor to the red, green, or purple belief.
The sixth and seventh condition correspond to a tie from the light blue actor to that same belief.
The eighth condition checks if the latter two ties share the same agreement pattern.
The last two conditions ensure that the three statements forming the four-cycle are not the same.
If all eight conditions are true, the cubic root of the product of the temporal weights for the three past statement events is added to the statistic.
The cubic root balances out the multiplication of three values in the $[0, 1]$ interval.
The procedure is repeated for all combinations of past $e_i, e_j, e_k$ statement events.

We posit that bonding is a foundational element of endogenous coalition formation in policy debates, i.\,e., actors adopt beliefs from congruent other actors in their effort to exercise discursive power via coalition building.
We expect that this mechanism plays a role in actors' decisions to contribute statements to a debate:
\begin{hypothesis}
	The stronger the bonding incentive is from the perspective of actor $s_e$ at time $t_e$ with regard to policy belief $r_e$ and agreement pattern $a_e$, the more likely $s_e$ will adopt $r_e$ at time $t_e$ with stance $a_e$.
\end{hypothesis}
``Incentive'' means that the past structure of the network before $t_e$ forms a strong signal to $s_e$ to adopt $r_e$ at $t_e$ with agreement pattern $a_e$ if $s_e$ considers bonding as a viable strategy to decide on making statements.

\subsection{Learning through Bridging}
Bonding alone, however, would lead to a compartmentalization of the network over time, where two coalitions with one distinct policy belief each would prevail \citep{leifeld2014polarization, altafini2012dynamics}.
Reality is more complex:
Coalitions are composed of more than one policy belief; more than two coalitions can exist empirically; and there can be actors who do not associate perfectly with one coalition or another \citep{leifeld2013reconceptualizing}.

Moreover, bonding, while useful and plausible, does not discriminate between cases where many other actors stated the same few policy beliefs with the same agreement pattern in the past (i.\,e., many actors, few beliefs) and cases where few other actors stated many beliefs with the same agreement pattern in the past (i.\,e., few actors, many beliefs).
For example, an actor $s_1$ may be particularly inclined to learn belief $r_1$ from another actor $s_2$ if the two actors exhibited many congruent policy beliefs before.
This would indicate a strong case of bonding and trust between these two actors, as hypothesized in the previous case.
Conversely, an actor $s_1$ may be inclined to learn a policy belief stance by observing that many other actors $s_i$ who had at least minimal agreement with $s_1$ in the past---even if otherwise not particularly close---agreed with the same stance as the focal actor $s_1$.
This would indicate that the actor takes cues not from a few highly trusted sources but from many moderately trusted sources.
When many minimally compatible actors have stated a policy belief before, actors may reach out of their immediate circle and adopt information from these ideologically relatively remote peers because many weak signals combine into a strong signal.

We control for bridging in a separate statistic because the bonding statistic otherwise contains elements of bridging as well.
Including both the bonding and bridging statistic then separates the many-actors--few-beliefs case of bridging from the few-actors--many-beliefs case of bonding.

This bridging behavior in policy debates is akin to the one documented in policy networks, where actors form collaborative ties outside of their neighborhood to seek out new information \citep{berardo2010self, berardo2014bridging, berardo2016understanding, henry2011belief}.

The distinction between bonding and bridging in policy debates also embodies the general dilemma between exploitation and exploration in decision situations \citep{cohen2007should}.
For instance, animals face the decision between exploiting known food sources with lower yield and known nutritional value and exploring potentially dangerous and low-yield but potentially diverse and high-yield new pathways in their food search \citep{hills2015exploration}.
In policy debates, actors similarly face the choice between re-iterating known positions from within their coalition (bonding or exploitation) and diversifying their policy belief portfolio by foraging for new ideas that are compatible with their existing beliefs outside their own coalition or in more remote parts of their coalition, perhaps another sub-coalition (bridging or exploration).

Bridging is an important building block of coalition formation and maintenance because it introduces innovation into coalitions.
The ACF argues that learning takes place predominantly within coalitions, but to a certain extent also between coalitions, in order to ensure that coalitions adjust to innovative information or solutions.
For example, if an external event, such as the publication of a significant scientific report or a natural catastrophe, introduces new beliefs to some actors, the bridging mechanism serves to diffuse these innovations to other actors, even in different coalitions.

However, actors will only adopt these policy beliefs if they are held by credible sources.
Actors only adopt policy beliefs from other actors if these other actors have a minimum degree of belief congruence with the focal actor.
They would not want to learn from their nemesis.
The bridging statistic captures this kind of innovation learning from minimally congruent source actors by assessing the share of other actors who both agreed with the focal belief stance and shared at least one other belief stance in the past among those who referred to the belief (with a compatible or incompatible agreement pattern) in the past.

The bridging statistic has three parts.
The first one is an indicator function that measures if an actor $s$ has had any previous agreement with focal actor $s_e$.
\begin{equation}
	\begin{split}
		\delta(s, e, E_{t - 1}) = \Bigg[ \Bigg( \sum_{e_i \in E_{t - 1}} \sum_{e_j \in E_{t - 1}} & [s_{e_i} = s] [s_e = s_{e_j}] [r_{e_i} = r_{e_j}] \\
		& [a_{e_i} = a_{e_j}] [r_e \neq r_{e_i}] [t_{e_i} < t_{e_j}] \Bigg) > 0 \Bigg]
	\end{split}
\end{equation}
It sums over any two statements in the past event sequence, $e_i$ and $e_j$, which must not be identical as per the last condition.
The first two conditions ensure that only statements are considered for $e_i$ that correspond to the focal actor $s$ and for $e_j$ that correspond to the current statement actor $s_e$, in order to count in how many past instances these two actors had shared agreement on beliefs.
The third and fourth condition check for identical belief--agreement combinations in this count.
If the overall count is positive, the indicator function yields 1.

The second part captures the amount of recent agreement over the current focal belief $r_e$ by the other actor $s$ with respect to focal actor $s_e$:
\begin{equation}
	\gamma(s, e, E_{t - 1}) = \sum_{e_i \in E_{t - 1}} \left( [s = s_{e_i}] [r_e = r_{e_i}] [a_e = a_{e_i}] w\left(t_{e_i}, t_e, T_{1/2}\right) \right)
\end{equation}
The three conditions indicate whether a previous statement $e_i$ was made by actor $s$ about the same belief and with the same agreement pattern as the focal statement event $e$ that is being considered.
If so, a temporally decayed weight is added to the count.

The third part is the (temporally weighted) number of actors who have referred to the focal belief $r_e$, whether in agreement or disagreement. 
It serves to standardize the bridging statistic by expressing the share of minimally compatible actors who agreed with the focal belief over all actors who referred to the belief.
\begin{equation}
	\kappa(s, e, E_{t - 1}) = \sum_{s \in S \setminus s_e} \sum_{e_i \in E_{t - 1}} \left( [s = s_{e_i}] [r_e = r_{e_i}] w\left(t_{e_i}, t_e, T_{1/2}\right) \right)
\end{equation}

Finally, the bridging statistic is defined as the sum of the product of the first two functions over all eligible actors as per the third function.
\begin{equation}
	u_\text{bridging}(e, E_{t-1}, T_{1/2}) = \frac{\sum_{s \in S \setminus s_e} \left( \delta(s, e, E_{t - 1}) \times \gamma(s, e, E_{t - 1}) \right)}{\kappa(s, e, E_{t - 1})}.
\end{equation}
We expect that the bridging mechanism contributes strongly to the topology of the policy debate:
\begin{hypothesis}
	The stronger the bridging incentive is from the perspective of actor $s_e$ at time $t_e$ with regard to policy belief $r_e$ and agreement $a_e$, the more likely $s_e$ will adopt $r_e$ at time $t_e$ with stance $a_e$.
\end{hypothesis}

\subsection{Learning through Repulsion}
Coalitions, or groups more generally, form not just through positive imitation, but also by repulsion between different coalitions or groups \citep{skvoretz2013diversity}.
For example, the literature on advocacy coalitions stresses how members of coalitions distrust opponents in policy processes \citep{fischer2016dealing}.
Once coalitions become entrenched via bonding and bridging, they stress dissimilarities with other coalitions.

Repulsion is the negative equivalent of bonding.
Rather than imitating those who have repeatedly displayed congruent stances (= bonding), actors oppose those who have repeatedly displayed opposite stances (= repulsion).
At the micro level, while bonding is operationalized through positively balanced four-cycles, or positive reciprocity, repulsion is operationalized through negatively balanced four-cycles along the event sequence, or negative reciprocity.

Repulsion means that disagreements between two actors are reinforced by the perception of being in different coalitions.
If actors $s_1$ and $s_2$ had previous disagreements over policy beliefs $r_1$, then it is likely that these conflicts are carried forward through new disagreements over other policy beliefs.
For example, if actor $s_1$ stated policy belief $r_1$ in a positive way while actor $s_2$ stated $r_1$ in a negative way and if $s_2$ stated belief $r_2$ in a negative way as well, then $s_1$ will tend to reinforce the interpersonal or inter-organizational conflict by stating $r_2$ in a positive way (third row in Figure~\ref{fig:mechanisms}).
Repulsion is defined in a similar way as bonding, just with opposing, rather than congruent, stances:
\begin{equation}
	\begin{split}
		u_\text{repulsion}(e, E_{t-1}, T_{1/2}) = & \sum_{e_i \in E_{t-1}} \sum_{e_j \in E_{t-1}} \sum_{e_k \in E_{t-1}} \bigg\{ [s_{e_i} \neq s_e] [r_{e_i} = r_e] [a_{e_i} \neq a_e] [s_{e_j} = s_e] \\
		& [r_{e_j} \neq r_e] [s_{e_k} = s_e] [r_{e_k} \neq r_e] [a_{e_j} \neq a_{e_k}] [t_{e_i} < t_{e_j}] [t_{e_j} < t_{e_k}] \\
		& \sqrt[3]{ w(t_{e_i}, t_e, T_{1/2}) w(t_{e_j}, t_e, T_{1/2}) w(t_{e_k}, t_e, T_{1/2}) } \bigg\},
	\end{split}
\end{equation}
The differences relative to bonding are the eighth condition, where past co-referrals to the same belief are conflictual, rather than congruent, and the third condition, where these past conflicts translate into a new conflict such that focal actor $s_e$ chooses the opposite stance of the other actor $s_{e_i}$ again.

Repulsion can be interpreted from an actor perspective or from an issue conflict perspective.
In the former view, actors learn from past conflicts and extend their existing conflicts to other issues.
In the latter view, there can be fundamental differences between sets of issues; for instance, policies $r_1$ and $r_2$ may be intrinsically linked because they belong to some coherent ideology $I_1$, and, similarly, policies $r_3$ and $r_4$ may be linked by some other overarching ideology $I_2$.
In that case, it is natural for actors adhering to ideology $I_1$ to support $r_1$ and $r_2$ but to reject $r_3$ and $r_4$.
In this view, repulsion is an intrinsic function of what the Advocacy Coalition Framework calls ``deep core beliefs'' \citep{sabatier1993policy}.

However, both perspectives can be reconciled:
Even if these fundamental predetermined cleavage lines exist across coalitions, actors will still \emph{learn} from the past behavior of their opponents as well as their overall position in the network what their stance on policy belief $r_e$ should be, given the overarching ideologies.
For example, if $s_1$ has a positive view on $r_2$ and recognizes $s_2$'s opposing, negative position on $r_2$, then $s_1$ can learn from $s_2$'s negative recent mention of $r_1$ that $r_2$ and $r_1$ are in the same compatible ideological package and that they should mention it in a positive way to maintain ideological congruence, or structural balance, thereby distancing themselves further from $s_2$.

In line with these considerations along the lines of structural balance and learning, we expect repulsion between actors to play a strong role in shaping the topology of the policy debate through reinforcement of differences between coalitions:
\begin{hypothesis}
	The stronger the incentives for repulsion are from the perspective of actor $s_e$ at time $t_e$ with regard to policy belief $r_e$ and agreement pattern $a_e$, the more likely will $s_e$ adopt $r_e$ at time $t_e$ with stance $a_e$.
\end{hypothesis}

\section{Data Collection}
We test our theory of endogenous coalition formation in policy debates in an empirical case study.
Using the software Discourse Network Analyzer (\url{https://www.github.com/leifeld/dna}, accessed 23 May 2024), we content-analyzed 1,842 newspaper articles from Frankfurter Allgemeine Zeitung, a German news outlet with center--right news coverage of politics, about the policy debate on the German pension system.
Articles were selected using a general search phrase (\texttt{*rente*}, the German word for pension), and false positives were manually removed from the set of documents.
6,704 statements of 245 political actors over nine years (1993--2001) were coded by two expert coders, and inter-coder reliability was checked post-hoc by two separate coders and was found to be consistently high.
The coding scheme consisted of four variables: the person who made a statement, the organization the person belonged to, the policy belief stated by the person or organization, and a dummy variable indicating support or rejection of the belief by the actor.
In the analysis, organizations were used as actors, in line with previous analyses of policy networks \citep{berardo2010self, laumann1987organizational, leifeld2012information}.
A total of 69 beliefs were coded.
These were policy instruments that reflected normative stances on what to do about the imminent financing problem of the pension system.
Examples include ``install fertility incentives to create more contributors,'' ``incentivize private savings to complement the state pension,'' or ``increase pension age,'' among others.

\begin{figure}
	\centering
	\includegraphics[width=1.0\textwidth]{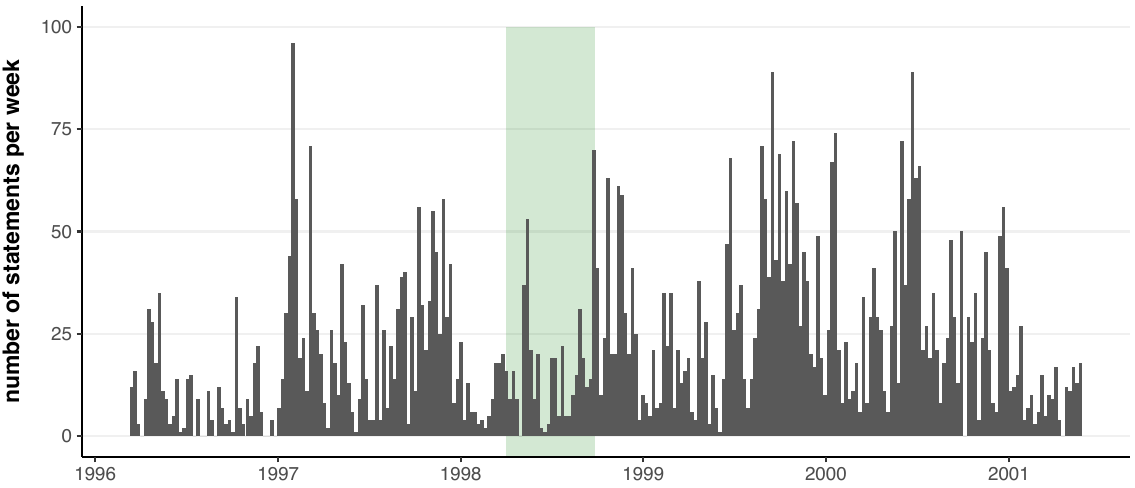}
	\caption{Frequency of statements per week. Election campaign highlighted.}
	\label{fig:frequencies}
\end{figure}

The start of the observation period was January 1993.
In May 2001, a major policy reform introduced privatization measures, which marks the end of the observation period.
We use the first 657 statements from January 1993 to March 1996 (200 event days) to calibrate the endogenous network statistics. 
With a half-life parameter of 30 event days, the network of past events is sufficiently large for all events issued after March 1996 to yield unbiased (``burnt-in'') network statistics.
Based on these considerations, we define March 13, 1996 as the start of the analysis period.
This time period also covers the national election at the end of September 1998 and the election campaign starting in March 1998.
The pension debate was one of the key topics in the election campaign.
Figure~\ref{fig:frequencies} shows the frequency of statement events per week over the analysis period, with the shaded area representing the duration of the election campaign.

The policy debate on the German pension system in the wake of population ageing is an ideal case study for several reasons.
First, Germany does not have a journalistic norm of balanced reporting, which would bias the measurement of coalition formation through newspaper data \citep{boykoff2004balance}.
We acknowledge that the process by which an actor gets to speak in the newspaper is still affected by agenda setting through the media, but actors have a high degree of autonomy in what they say, and we can control for activity and popularity effects in the statistical model.
Second, the debate featured a high amount of uncertainty among actors over how to solve the policy problem.
The debate is situated in a historical setting where population ageing was, for the first time, put on the national political agenda.
This ensures a lively debate, where a number of very different arguments and positions are exchanged.
Third, the topic of public pensions is broad enough that the debate is multi-dimensional rather than two-dimensional and that more than two coalitions can be observed at some time points \citep{leifeld2013reconceptualizing, leifeld2016policy}.

\section{Methods}
We created an event sequence containing 38,678,640 possible events (245 actors $\times$ 69 beliefs $\times$ 2 stances $\times$ 1144 unique event days) as rows in a rectangular dataset, defining the probability space with all possible combinations over the four event variables actors, beliefs, agreement, and time.
In addition to the values of the four variables, we denoted as 1 which events occurred (``true events'') and as 0 which events did not occur (``null events''), forming the dependent variable.

The time variable, which indicates the day the statement event was reported (and the corresponding null events that could have been reported), acts as a stratum or grouping variable to denote which events and null events belong together into a risk set of which events happened or could have happened on a day.

We computed the three network statistics for bonding, bridging, and repulsion for every event and every null event given the prior event sequence, taking into account only the true events in the calculation of the statistics.
The three temporally changing statistics were added to the dataset as columns.

We furthermore added a dummy variable indicating whether the respective actor is a government actor, such as a federal ministry or agency.
We expect these actors to behave differently and act in a neutral way because they have an official mandate.
To test for these differences in behavior between government actors and other actors, we included interaction effects between the government variable (``GOV'') and the three network statistics bonding, bridging, and repulsion in a separate model specification.

The interaction should have a negative sign as policy belief learning should be less pronounced in government actors than in other types of actors.
The main effect for this dummy variable should be positive because government actors are often in charge of issuing statements about policy making. Therefore, they should have a relative high statement density.

Three additional statistics commonly used in applications of relational event models \citep{brandes2009networks, butts2008relational, lerner2013modeling} were added as additional columns and served as control variables: 
First, \emph{inertia} captures the tendency for recent events with the same variable values to be replicated along the event sequence (similar to the effect of lagged variables in time series contexts).
This corrects for duplicates in the dataset due to multiple reporting of the same statement.
Second, \emph{belief popularity} captures the tendency for recently frequently stated beliefs to become more likely.
This serves to control for issue attention cycles \citep{downs1972up}, i.\,e., swings in the number of times a specific belief was recently stated.
Third, \emph{actor activity} captures the tendency for frequently active actors to make yet more new statements.
It models activity differentials in a similar way as a random effect such that less active actors are modeled to have a lower statement probability and more active actors are modeled to have a higher statement probability.

With this dataset, we estimated a stratified Cox proportional hazard model using the entries of the time variable as strata.
As each stratum contains identical risk sets of actor--belief--agreement combinations, the stratified Cox model reduces to a conditional logit model, which is common in choice analysis and case-control studies and is easy to implement and estimate in standard statistical software with common add-on packages.
In this setup, the network statistics bonding, bridging, repulsion, inertia, belief popularity, and actor activity along with the exogenous covariate GOV serve to explain the occurrence of each new event in the sequence relative to all the events that could have occurred (the null events).

The use of network statistics to capture dependencies between observations in a survival model is known as a relational event model (REM) and was first proposed by \citet{butts2008relational}.
By including the bonding, bridging, and repulsion statistics, we extend REMs accommodate custom network statistics formed over a signed two-mode network.
Previous research predicted the sign of an interaction in a network event sequence as a function of network statistics formed over the past event sequence \citep{denooy2013polarization, lerner2013modeling}.
In this application, tie location (actors and beliefs) and sign (agreement) are explained jointly in a single model predicting true versus null events.

The event sequence is not perfectly ordered because several events per day were sometimes recorded.
Efron's tie breaker approximation was employed to mitigate any bias resulting from this measurement error caused by binning into days \citep{efron1977efficiency}.

Coefficients can be interpreted as changes in log-odds ratios of turning a null event into an event if the independent variable increases by one unit.
Given the complex nature of the network statistics, a one-unit change can be caused by different things, including changes in the timing of any past event involved and the number of times the statistic is present in the past event sequence.
The coefficients were rescaled to permit more comparable effect sizes, without changing their significance (see Table \ref{tab_rescale} in the Appendix for exact rescale values and interpretations thereof). 
For the purposes of theory development and testing, it suffice to interpret the direction and statistical significance along with the substantive meaning below.

\section{Results}
\begin{figure}
	\centering
	\includegraphics[width=1.0\textwidth]{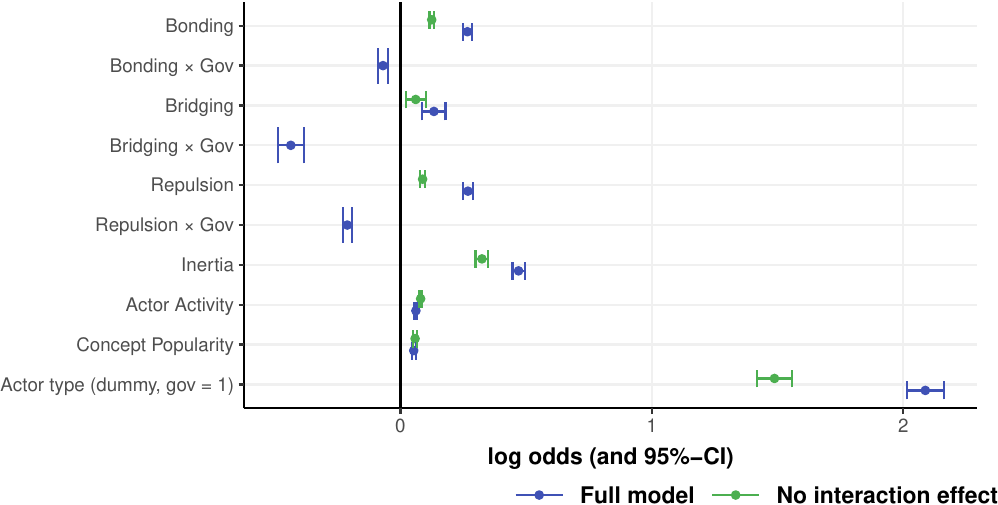}
	\caption{Estimates from the stratified Cox model. The basic (main) model in green; model with GOV interactions in blue. Bonding, bridging, and repulsion all have positive and significant coefficients. Note: Table \ref{tab_rem} in the Appendix reports the accompanying regression table.}
	\label{fig:coefficients}
\end{figure}

The statistical analysis supports the role of bonding, bridging, and repulsion in the emergence of new statements in this policy debate.
Figure~\ref{fig:coefficients} shows a visual representation of the estimates for the main model (green) and the model with interactions between government actors and the three main variables.
All three effects are positive and significant and allow us to reject the null hypotheses that these endogenous mechanisms played no roles in how statements were made in the debate.
As part of the same process, political actors turn to trusted coalition members in an effort to exploit the local coalition neighborhood for inspiration on which beliefs to contribute to the debate (= bonding), explore the wider network topology even across coalitions to learn policy beliefs using the wisdom of the crowd (bridging), and learn from past conflicts how to position themselves in opposition to earlier political opponents (repulsion).

Bonding establishes, and reinforces the ideological congruence within coalitions.
Repulsion leads to continued polarization between these coalitions.
Bridging counterbalances repulsion and bonding by introducing weak learning ties and belief overlap between coalitions.
Together, these mechanisms produce discourse networks with polarized but not fully disjoint coalitions, which exist in a steady-state equilibrium during which they can sometimes disband and sometimes grow more together but on average form a loosely coupled system of distinguishable belief coalitions, in line with prior empirical findings at the macro level \citep[e.\,g.,][]{leifeld2013reconceptualizing} and prior theoretical simulation predictions \citep{leifeld2014polarization}.

Government actors display more activity than other actors, as per the last row in Figure~\ref{fig:coefficients}.
Verbose actors include the Ministry of Labor and Social Affairs (BMAS) and the large parties.
We find in the interaction model that government actors display lower levels of bonding, bridging, and repulsion than other actor types, such as banks and insurance companies, interest groups, trade unions and charities, and industry and employers' associations, with negative interaction effects between the government dummy and the network statistics and positive main effects.

\begin{figure}
	\centering
	\includegraphics[width=0.5\textwidth]{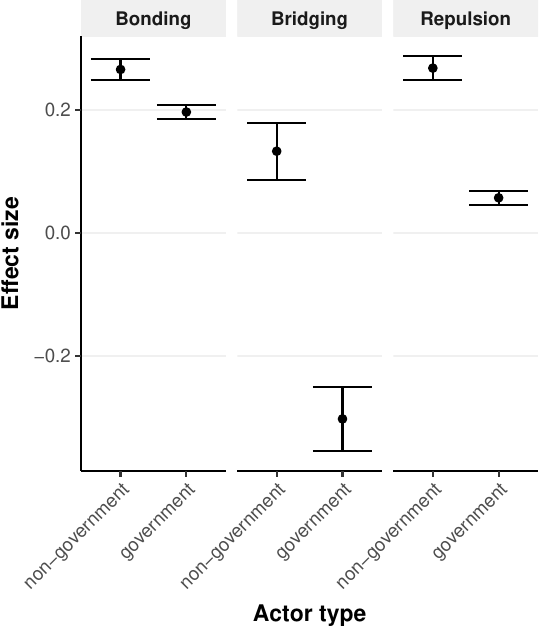}
	\caption{Marginal effects for interaction terms.}
	\label{fig:marginal}
\end{figure}

The marginal effects plot in Figure~\ref{fig:marginal} facilitates the interpretation of these interaction effects on the probability scale.
Government actors are less prone to bonding and repulsion than other actors, which makes sense because interest groups should be more vested in the causes of their coalitions than government actors.
In the German neo-corporatist model, the state takes a mediating role between employers and trade unions and should not be captured by interest group coalitions.
Bridging has a positive effect for most actors, but government actors actively ``burn bridges'' by systematically ignoring policy beliefs from the wider discourse network, as shown by the negative coefficient.
For government actors, the presence of bridging opportunities makes them less likely to use them.
Our ad-hoc explanation is that it would be detrimental for government actors to be captured, and since the remaining network is composed mostly of vested interests, adopting their beliefs, even if there is minimal belief overlap, would be detrimental to the credibility and neutrality of the government and leading parties.
Maintaining a facilitating and mediating role may have been paramount as this particular debate was characterized by fiercely polarized camps of banks and insurance companies with their interest groups and employers' and industry associations on one side vis-à-vis trade unions and social interest groups on the other side \citep{leifeld2013reconceptualizing}.

\begin{figure}
	\centering
	\includegraphics[width=0.8\textwidth]{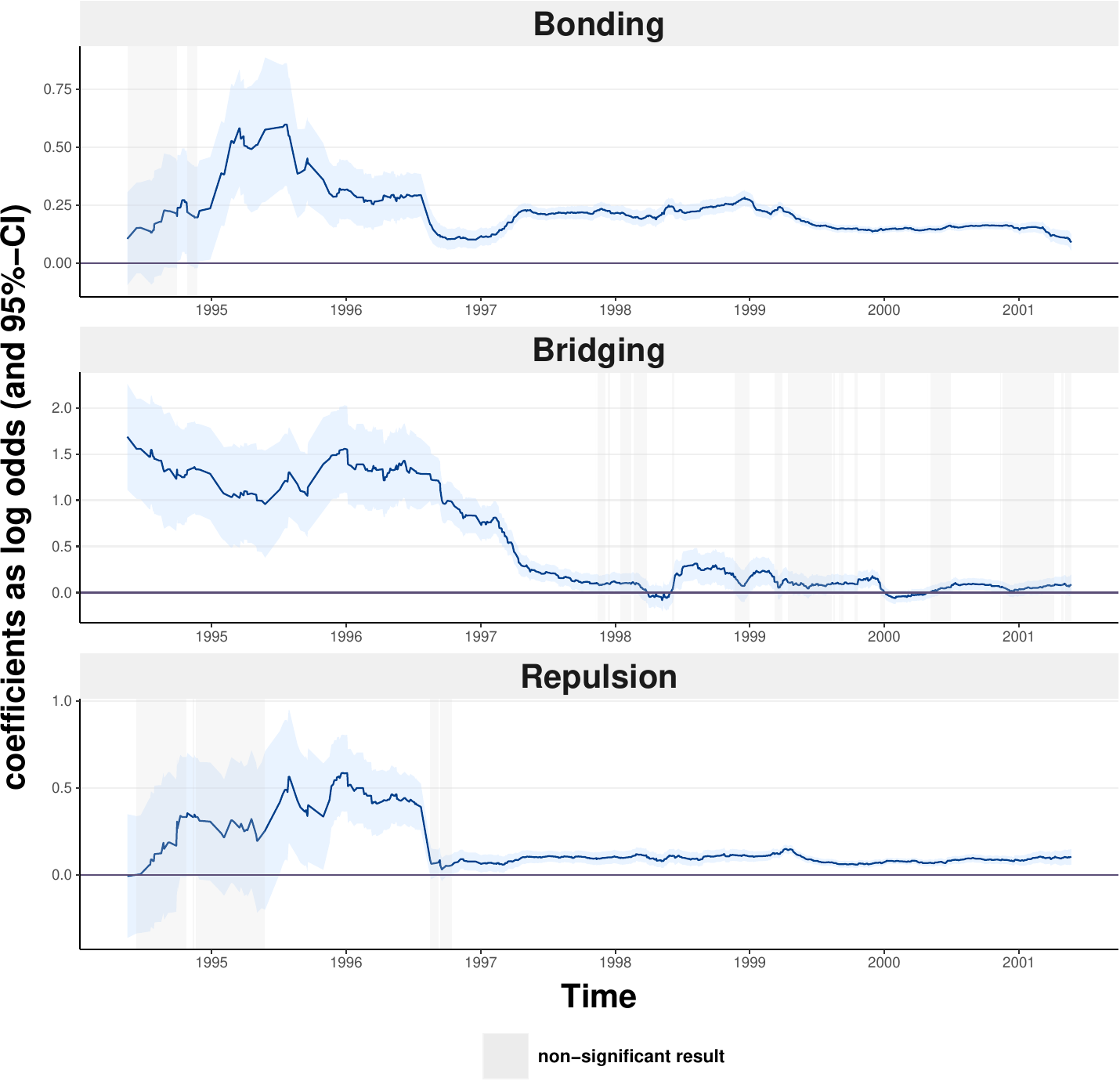}
	\caption{Assessment of temporal heterogeneity in coefficients via repeated re-estimation in moving time windows of one year ($\pm 182$ days) around a given date. Blue lines indicate coefficients; blue-shaded polygons represent 95\,\% confidence intervals; gray-shaded areas are non-significant time periods (but mind the lower statistical power in one-year time slices).}
	\label{fig:time}
\end{figure}

An important aspect in modeling a temporally emerging complex system is parameter homogeneity.
As one single parameter was estimated per variable, the parameters should not switch between positive and negative or show very clear trends over time.
Figure~\ref{fig:time} assesses this question using re-estimated models within one-year moving-average time windows.
The coefficients remain in the positive range with very few exceptions.
At most time points, they are significant from zero.
The bonding and repulsion mechanisms, which are responsible for establishing strong, congruent coalitions, show some non-significant results before 1997, and the bridging mechanism, which counteracts polarization, shows some non-significant results starting in 1998.
This is in good keeping with the case study details because the policy debate became polarized shortly before the election campaign in 1998.
During this very polarized time (1998--2001), bridging played a lesser role, while bonding and repulsion were not consistently observed during pre-polarization stages of the debate \citep{leifeld2013reconceptualizing}.
While the effect sizes in the first two years were larger, so were the confidence intervals due to the sparsity of observations.

\begin{figure}
	\centering
	\includegraphics[width=1.0\textwidth]{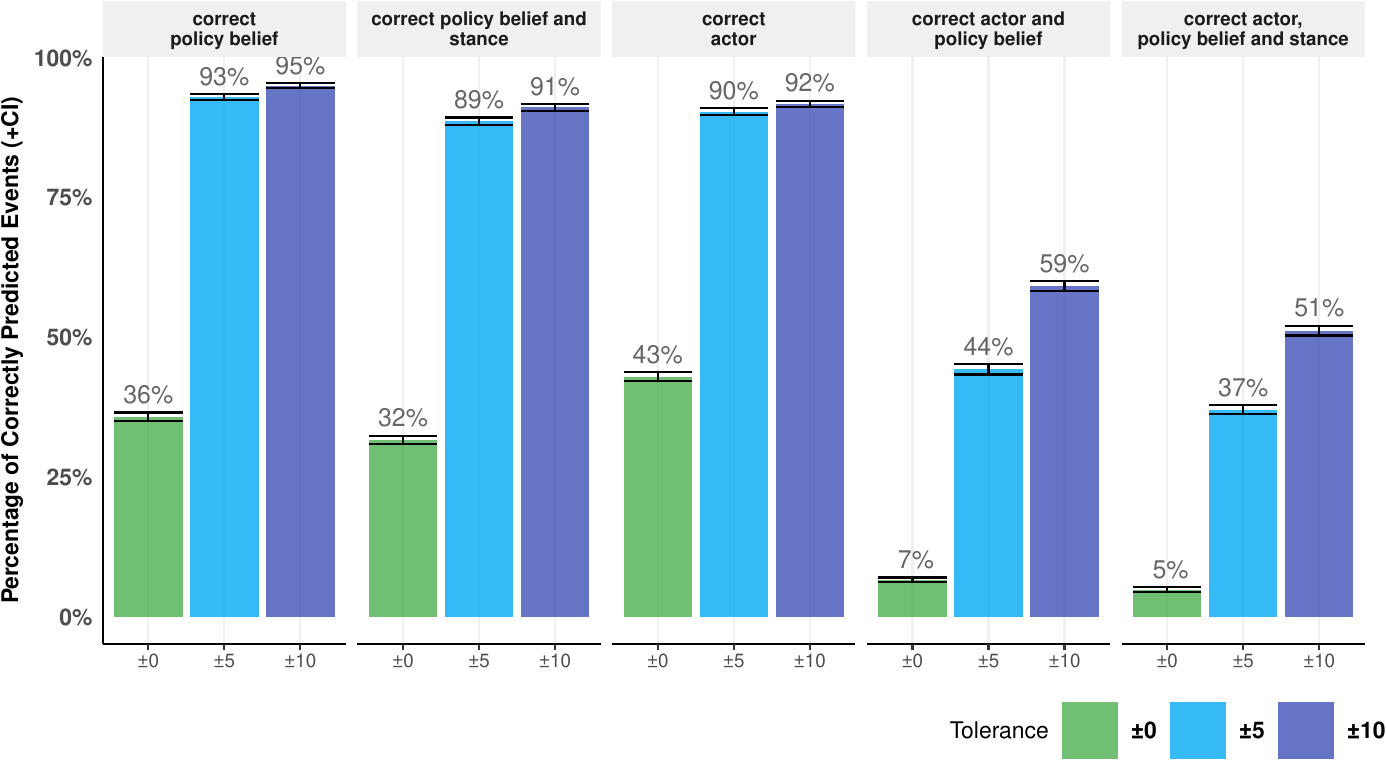}
	\caption{Out-of-sample prediction performance of new events with 100 simulated sequences. With more demanding criteria, the predictive fit becomes worse.}
	\label{fig:prediction}
\end{figure}

\begin{figure}
	\centering
	\includegraphics[width=1.0\textwidth]{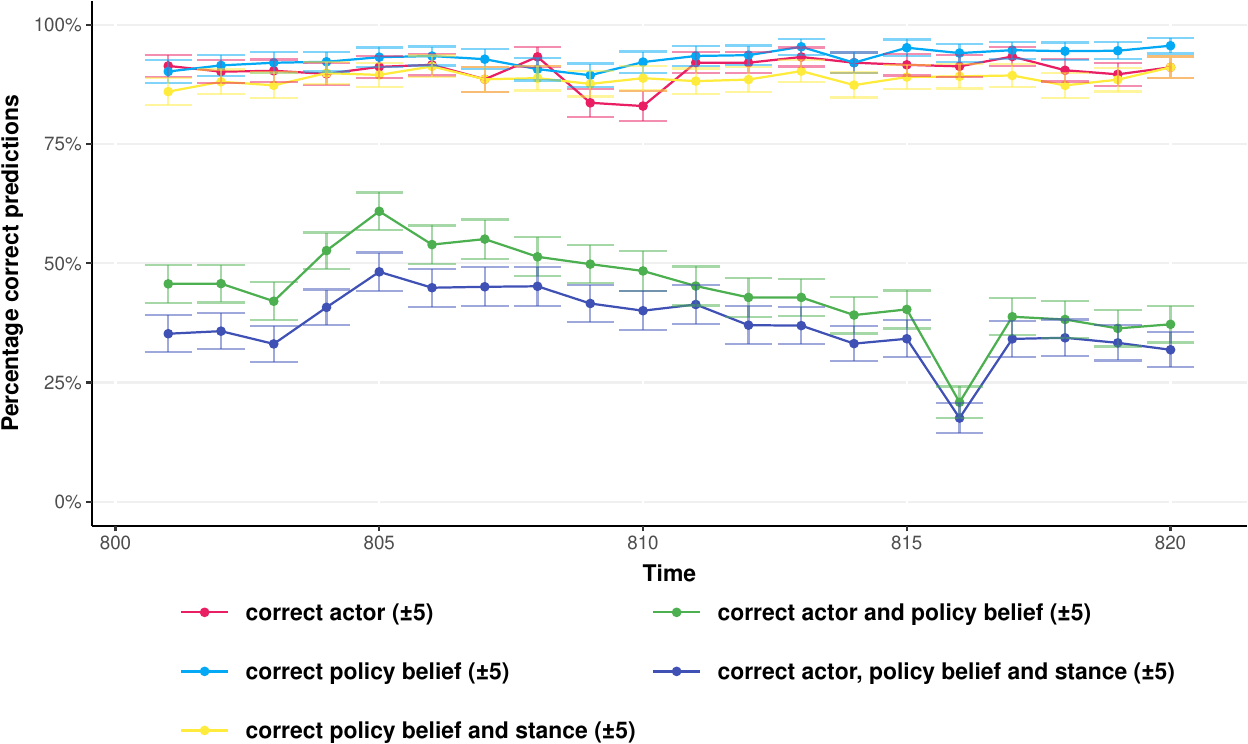}
	\caption{Quality of out-of-sample predictions over time. The harder prediction tasks decay slowly in quality because uncertainty increases.}
	\label{fig:predictiontime}
\end{figure}

To evaluate how well the model with the three endogenous building blocks of coalition formation works, we predicted new statements out of sample in Figures~\ref{fig:prediction} and~\ref{fig:predictiontime}.
Predicted probabilities are generated step by step for the next 20 events in the sequence, starting at event time 800 (out of 1144 unique event days). Each new predicted event is fed into the sequence as a basis for the recalculation of network statistics and updated coefficients and prediction of the next event. 
Complete actor--belief--agreement--time forecasts with zero time tolerance are correct in 5\,\% of the predictions, compared to the true sequence that unfolded in the observed data.
With five or ten days tolerance when the event occurs, the share goes up to 37\,\% or 51\,\%, respectively.
As criteria are further relaxed towards the left of the plot (e.\,g., predicting only policy belief and agreement with a five-day tolerance), the share of correct predictions exceeds 90\,\%.

The second prediction plot (Figure~\ref{fig:predictiontime}) unpacks the temporal development of the out-of-sample prediction quality further.
The simpler prediction tasks have no decay in the quality of predictions as time since the last observations progresses.
The more complex combinations, such as requiring correct actor, belief, agreement, and a tolerance of five days, see a decline in prediction accuracy over time.
The model is better than chance would predict, and future research needs to improve upon the theory presented here to become more accurate in forecasting political debates.

\section{Conclusion}
The results provide empirical evidence for the three mechanisms of learning that lead to endogenous coalition formation and maintenance.
This provides us with an understanding of how and why advocacy or discourse coalitions are formed and maintained and, more broadly, how policy debates operate.

The question of how political actors learn within and across coalitions and, thereby, form and maintain their coalitions or change the structure of a policy subsystem, has been central to the literature on advocacy coalitions since the 1980s, but little progress has been made towards developing micro-level theory.
The model presented here can be regarded as a first attempt to provide a general theory of policy learning in the context of coalition formation and maintenance, with applicability to advocacy and discourse coalitions and potentially discourse in more general contexts, such as online and social media discourse, deliberation in committees, or other settings.
Future research should evaluate how well the combination of bonding, bridging, and repulsion explain the course of debates in these areas.

Students of polarization may find new inspiration in this research to go beyond measuring polarization as a macro phenomenon in networked systems and towards modeling the generative process underlying polarized systems at the micro level.

The literature on policy process theory, which includes the ACF and other theories at the subsystem level, is in dire need of formal models to progress from ``frameworks'' to ``theories'', even if it may mean that some existing theoretical ideas turn out to be untenable or useless.
In this particular instance, we have provided a theory that addresses one of the most central unresolved holes in the ACF and perhaps policy process theory more widely.
Students of the politics around public policy should apply, test, and---most importantly---extend this theory of learning within and between coalitions.
If they succeed, their new developments may also improve the predictive performance of the model.
One could leverage such developments into a support tool for consulting elite actors on political strategy through simulation-based thought experiments of how an actor's contributions may change the course of the debate.

Finally, we would like to point out that the policy debate analyzed here was selected carefully, but may deviate from policy debates in other political systems, arenas, data sources, policy domains, or contexts more generally.
This opens the door to an exciting research area where debates in different contexts can be compared to the baseline findings established here.

\bibliographystyle{apalike}
\bibliography{manuscript}

\begin{thebibliography}{}

\bibitem[Altafini, 2012]{altafini2012dynamics}
Altafini, C. (2012).
\newblock Dynamics of opinion forming in structurally balanced social networks.
\newblock {\em PLOS ONE}, 7(6):e38135.

\bibitem[Antal et~al., 2005]{antal2005dynamics}
Antal, T., Krapivsky, P.~L., and Redner, S. (2005).
\newblock Dynamics of social balance on networks.
\newblock {\em Physical Review E}, 72(3):036121.

\bibitem[Berardo, 2014]{berardo2014bridging}
Berardo, R. (2014).
\newblock Bridging and bonding capital in two-mode collaboration networks.
\newblock {\em Policy Studies Journal}, 42(2):197--225.

\bibitem[Berardo and Lubell, 2016]{berardo2016understanding}
Berardo, R. and Lubell, M. (2016).
\newblock Understanding what shapes a polycentric governance system.
\newblock {\em Public Administration Review}, 76(5):738--751.

\bibitem[Berardo and Scholz, 2010]{berardo2010self}
Berardo, R. and Scholz, J.~T. (2010).
\newblock Self-organizing policy networks: Risk, partner selection, and
  cooperation in estuaries.
\newblock {\em American Journal of Political Science}, 54(3):632--649.

\bibitem[Boykoff and Boykoff, 2004]{boykoff2004balance}
Boykoff, M.~T. and Boykoff, J.~M. (2004).
\newblock Balance as bias: Global warming and the {US} prestige press.
\newblock {\em Global Environmental Change}, 14(2):125--136.

\bibitem[Brandes et~al., 2009]{brandes2009networks}
Brandes, U., Lerner, J., and Snijders, T. A.~B. (2009).
\newblock Networks evolving step by step: Statistical analysis of dyadic event
  data.
\newblock In {\em Social Network Analysis and Mining, ASONAM'09. International
  Conference on Advances in Social Network Analysis and Mining}, pages
  200--205. IEEE.

\bibitem[Butts, 2008]{butts2008relational}
Butts, C.~T. (2008).
\newblock A relational event framework for social action.
\newblock {\em Sociological Methodology}, 38(1):155--200.

\bibitem[Cartwright and Harary, 1956]{cartwright1956structural}
Cartwright, D. and Harary, F. (1956).
\newblock Structural balance: A generalization of {H}eider's theory.
\newblock {\em Psychological Review}, 63(5):277--293.

\bibitem[Cohen et~al., 2007]{cohen2007should}
Cohen, J.~D., McClure, S.~M., and Yu, A.~J. (2007).
\newblock Should i stay or should i go? {H}ow the human brain manages the
  trade-off between exploitation and exploration.
\newblock {\em Philosophical Transactions of the Royal Society B: Biological
  Sciences}, 362(1481):933--942.

\bibitem[de~Nooy and Kleinnijenhuis, 2013]{denooy2013polarization}
de~Nooy, W. and Kleinnijenhuis, J. (2013).
\newblock Polarization in the media during an election campaign: A dynamic
  network model predicting support and attack among political actors.
\newblock {\em Political Communication}, 30(1):117--138.

\bibitem[Downs, 1972]{downs1972up}
Downs, A. (1972).
\newblock Up and down with ecology: The issue attention cycle.
\newblock {\em Public Interest}, 28(1):38--50.

\bibitem[Efron, 1977]{efron1977efficiency}
Efron, B. (1977).
\newblock The efficiency of {C}ox's likelihood function for censored data.
\newblock {\em Journal of the American Statistical Association},
  72(359):557--565.

\bibitem[Fischer et~al., 2016]{fischer2016dealing}
Fischer, M., Ingold, K., Sciarini, P., and Varone, F. (2016).
\newblock Dealing with bad guys: Actor- and process-level determinants of the
  `devil shift' in policy making.
\newblock {\em Journal of Public Policy}, 36(2):309--334.

\bibitem[Hajer, 1995]{hajer1995politics}
Hajer, M.~A. (1995).
\newblock {\em The Politics of Environmental Discourse: Ecological
  Modernization and the Policy Process}.
\newblock Clarendon Press, Oxford.

\bibitem[Hall, 1993]{hall1993policy}
Hall, P.~A. (1993).
\newblock Policy paradigms, social learning, and the state: The case of
  economic policymaking in {B}ritain.
\newblock {\em Comparative Politics}, 25(3):275--296.

\bibitem[Henry et~al., 2011]{henry2011belief}
Henry, A.~D., Lubell, M., and McCoy, M. (2011).
\newblock Belief systems and social capital as drivers of policy network
  structure: The case of {C}alifornia regional planning.
\newblock {\em Journal of Public Administration Research and Theory},
  21(3):419--444.

\bibitem[Hills et~al., 2015]{hills2015exploration}
Hills, T.~T., Todd, P.~M., Lazer, D., Redish, A.~D., and Couzin, I.~D. (2015).
\newblock Exploration versus exploitation in space, mind, and society.
\newblock {\em Trends in Cognitive Sciences}, 19(1):46--54.

\bibitem[Hojnacki, 1997]{hojnacki1997interest}
Hojnacki, M. (1997).
\newblock Interest groups' decisions to join alliances or work alone.
\newblock {\em American Journal of Political Science}, 41(1):61--87.

\bibitem[Junk, 2019]{junk2019diversity}
Junk, W.~M. (2019).
\newblock When diversity works: The effects of coalition composition on the
  success of lobbying coalitions.
\newblock {\em American Journal of Political Science}, 63(3):660--674.

\bibitem[Lasswell, 1950]{lasswell1950politics}
Lasswell, H.~D. (1950).
\newblock {\em Politics: Who Gets What, When, How}.
\newblock P.~Smith, New York.

\bibitem[Laumann and Knoke, 1987]{laumann1987organizational}
Laumann, E.~O. and Knoke, D. (1987).
\newblock {\em The organizational state: Social choice in national policy
  domains}.
\newblock Univ of Wisconsin Press.

\bibitem[Leifeld, 2013]{leifeld2013reconceptualizing}
Leifeld, P. (2013).
\newblock Reconceptualizing major policy change in the advocacy coalition
  framework: A discourse network analysis of {G}erman pension politics.
\newblock {\em Policy Studies Journal}, 41(1):169--198.

\bibitem[Leifeld, 2014]{leifeld2014polarization}
Leifeld, P. (2014).
\newblock Polarization of coalitions in an agent-based model of political
  discourse.
\newblock {\em Computational Social Networks}, 1(1):7.

\bibitem[Leifeld, 2016]{leifeld2016policy}
Leifeld, P. (2016).
\newblock {\em Policy Debates as Dynamic Networks: {G}erman Pension Politics
  and Privatization Discourse}.
\newblock Campus, Frankfurt am Main.
\newblock Distributed through the University of Chicago Press.

\bibitem[Leifeld and Haunss, 2012]{leifeld2012political}
Leifeld, P. and Haunss, S. (2012).
\newblock Political discourse networks and the conflict over software patents
  in europe.
\newblock {\em European Journal of Political Research}, 51(3):382--409.

\bibitem[Leifeld and Schneider, 2012]{leifeld2012information}
Leifeld, P. and Schneider, V. (2012).
\newblock Information exchange in policy networks.
\newblock {\em American Journal of Political Science}, 56(3):731--744.

\bibitem[Lerner et~al., 2013]{lerner2013modeling}
Lerner, J., Bussmann, M., Snijders, T. A.~B., and Brandes, U. (2013).
\newblock Modeling frequency and type of interaction in event networks.
\newblock {\em Corvinus Journal of Sociology and Social Policy}, 4(1):3--32.

\bibitem[Marvel et~al., 2011]{marvel2011continuous}
Marvel, S.~A., Kleinberg, J., Kleinberg, R.~D., and Strogatz, S.~H. (2011).
\newblock Continuous-time model of structural balance.
\newblock {\em Proceedings of the National Academy of Sciences},
  108(5):1771--1776.

\bibitem[Moyson, 2017]{moyson2017cognition}
Moyson, S. (2017).
\newblock Cognition and policy change: The consistency of policy learning in
  the advocacy coalition framework.
\newblock {\em Policy and Society}, 36(2):320--344.

\bibitem[Neal, 2020]{neal2020sign}
Neal, Z.~P. (2020).
\newblock A sign of the times? {W}eak and strong polarization in the {U.S.}
  {C}ongress, 1973--2016.
\newblock {\em Social Networks}, 60:103--112.

\bibitem[Quattrociocchi et~al., 2014]{quattrociocchi2014opinion}
Quattrociocchi, W., Caldarelli, G., and Scala, A. (2014).
\newblock Opinion dynamics on interacting networks: Media competition and
  social influence.
\newblock {\em Scientific Reports}, 4(4938).

\bibitem[Sabatier and Jenkins-Smith, 1993]{sabatier1993policy}
Sabatier, P.~A. and Jenkins-Smith, H.~C. (1993).
\newblock {\em Policy Change and Learning: An Advocacy Coalition Approach}.
\newblock Westview Press, Boulder, CO.

\bibitem[Sabatier and Weible, 2014]{sabatier2014theories}
Sabatier, P.~A. and Weible, C. (2014).
\newblock {\em Theories of the Policy Process}.
\newblock Westview Press, Boulder, CO.

\bibitem[Skvoretz, 2013]{skvoretz2013diversity}
Skvoretz, J. (2013).
\newblock Diversity, integration, and social ties: Attraction versus repulsion
  as drivers of intra-and intergroup relations.
\newblock {\em American Journal of Sociology}, 119(2):486--517.

\bibitem[Weible et~al., 2023]{weible2023advocacy}
Weible, C.~M., Olofsson, K.~L., and Heikkila, T. (2023).
\newblock Advocacy coalitions, beliefs, and learning: An analysis of stability,
  change, and reinforcement.
\newblock {\em Policy Studies Journal}, 51(1):209--229.

\bibitem[Weible et~al., 2009]{weible2009themes}
Weible, C.~M., Sabatier, P.~A., and McQueen, K. (2009).
\newblock Themes and variations: Taking stock of the advocacy coalition
  framework.
\newblock {\em Policy Studies Journal}, 37(1):121--140.

\bibitem[Zafonte and Sabatier, 1998]{zafonte1998shared}
Zafonte, M. and Sabatier, P.~A. (1998).
\newblock Shared beliefs and imposed interdependencies as determinants of ally
  networks in overlapping subsystems.
\newblock {\em Journal of Theoretical Politics}, 10(4):473--505.

\end{thebibliography}

\newpage

\appendix

\section*{Appendix}

\begin{figure}[h!]
	\centering
	\includegraphics[width=0.6\textwidth]{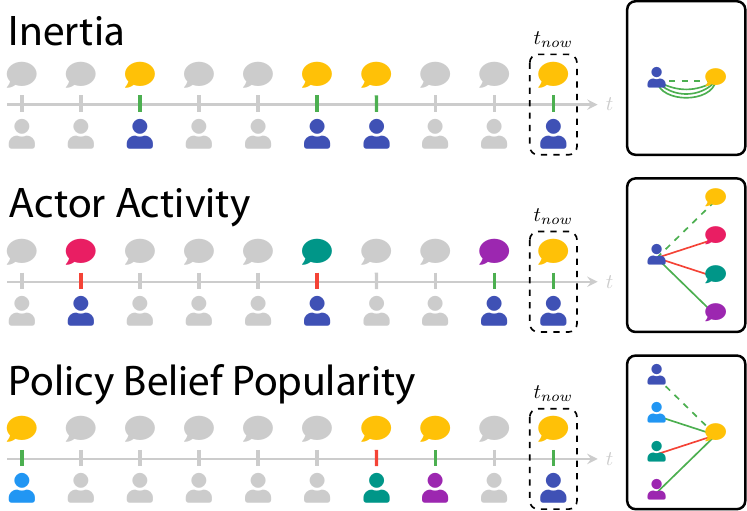}
	\caption{Illustration of network statistics used as control variables.}
	\label{fig:control}
\end{figure}

\renewcommand{\arraystretch}{1.2}
\begin{table}
\centering
\begin{tabular}{llp{9cm}}
	\toprule
	Variable & Value & Possible interpretation for the rescale value \\
	\midrule
    Bonding         & 7.578583      & $+10$ past bonding events (occurring 1, 3 and 10 days ago) \\
    Bridging        & 7.937005      & $+10$ past bridging events (with the focal event occurring 10 days ago)\\
    Repulsion       & 7.578583      & $+10$ past repulsion events (occurring 1, 3 and 10 days ago)\\
    Inertia         & 2.189874   & $+3$ past events (involving the same actor, same belief and same stance), occurring 7, 14 and 21 days ago.\\
    Activity        & 2.189874   & $+3$ past events (involving the same actor), occurring 7, 14 and 21 days ago.\\
    Popularity      & 2.189874   & $+3$ past events (involving the same belief), occurring 7, 14 and 21 days ago.\\
	\bottomrule	
\end{tabular}
\caption{Rescaling of endogenous network statistics. Possible interpretation of $\pm1$ unit increase. Note that other combinations of events (both quantity and temporal lags) can result in the same rescale values.}\label{tab_rescale}
\end{table}
\renewcommand{\arraystretch}{1}

\begin{table}[h!]
	\centering
	\begin{tabular}{l D{)}{)}{9)3} D{)}{)}{8)3}}
		\hline
		& \multicolumn{1}{c}{(1)} & \multicolumn{1}{c}{(2)} \\
		\hline
		Main hypotheses                          &                       &                      \\
		\quad Bonding (positive reciprocity)     & 0.27 \; (0.01)^{***}  & 0.12 \; (0.00)^{***} \\
		\quad Bonding $\times$ government org.   & -0.07 \; (0.01)^{***} &                      \\
		\quad Bridging                           & 0.13 \; (0.02)^{***}  & 0.06 \; (0.02)^{**}  \\
		\quad Bridging $\times$ government org.  & -0.44 \; (0.03)^{***} &                      \\
		\quad Repulsion (negative reciprocity)   & 0.27 \; (0.01)^{***}  & 0.09 \; (0.01)^{***} \\
		\quad Repulsion $\times$ government org. & -0.21 \; (0.01)^{***} &                      \\
		Controls                                 &                       &                      \\
		\quad Inertia                            & 0.47 \; (0.01)^{***}  & 0.32 \; (0.01)^{***} \\
		\quad Actor activity                     & 0.06 \; (0.00)^{***}  & 0.08 \; (0.00)^{***} \\
		\quad Belief concept popularity          & 0.05 \; (0.00)^{***}  & 0.06 \; (0.00)^{***} \\
		\quad Government (dummy)                 & 2.09 \; (0.04)^{***}  & 1.49 \; (0.04)^{***} \\
		\hline
		AIC                                      & 96255.40              & 98478.32             \\
		McFadden pseudo-R$^2$                    & 0.236                  & 0.219                 \\
        Num. events                              & 6043                  & 6043                 \\
		Num. obs.                                & 31916640              & 31916640             \\
		\hline
		\multicolumn{3}{l}{\tiny{$^{***}p<0.001$; $^{**}p<0.01$; $^{*}p<0.05$}}
	\end{tabular}
\caption{Results of the conditional logistic regression on the event rate}
\label{tab_rem}
\end{table}

\begin{figure}[h!]
	\centering
	\includegraphics[width=0.8\textwidth]{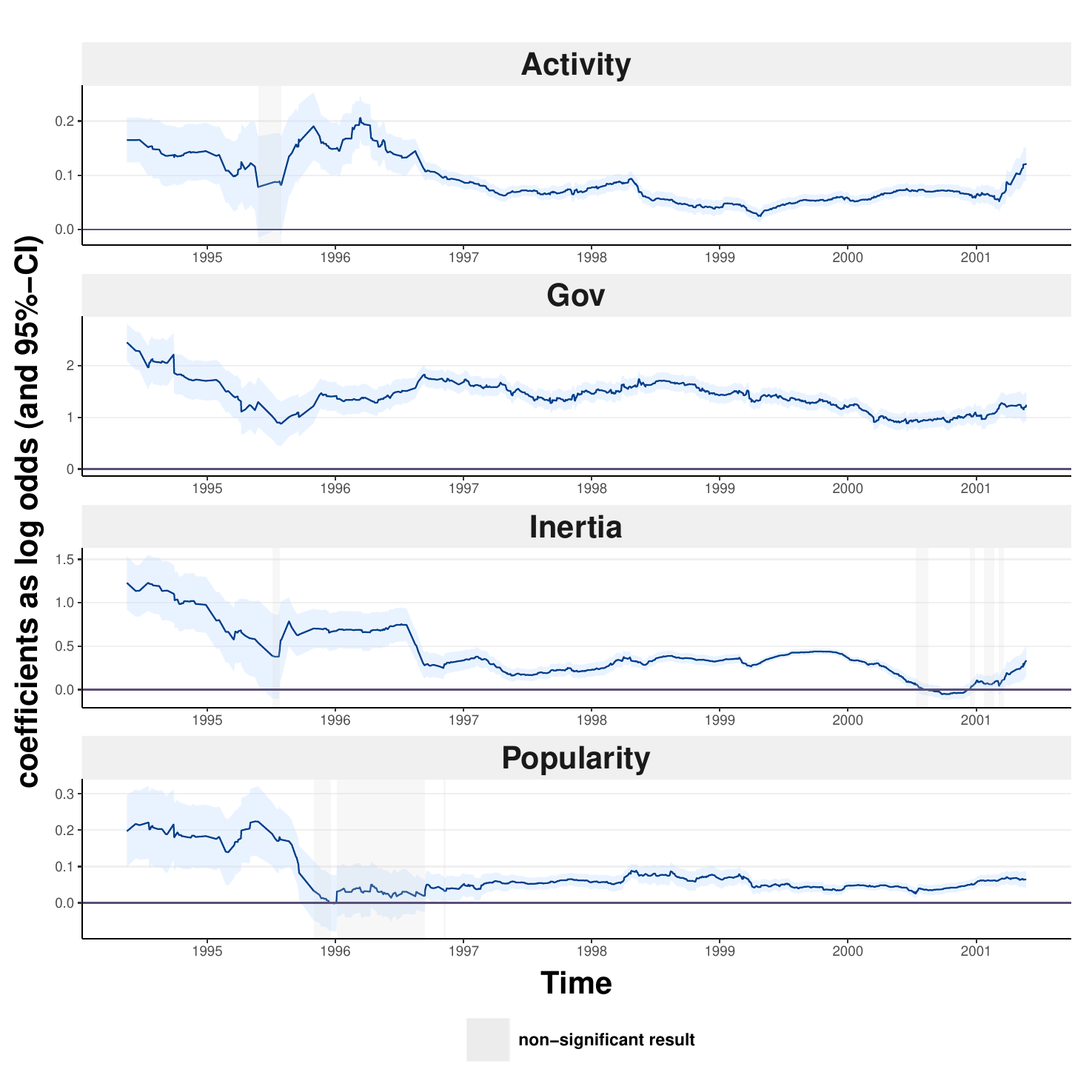}
	\caption{Assessment of temporal heterogeneity in coefficients for control variables.}
	\label{fig:controltemp}
\end{figure}

\end{document}